\documentclass[12pt,preprint]{aastex}
\begin{document}


\title{Binary Central Stars of Planetary Nebulae Discovered Through Photometric Variability V: The Central Stars of HaTr~7 and ESO~330-9}
\author{Todd C. Hillwig\altaffilmark{1}}
\altaffiltext{1}{Department of Physics and Astronomy, Valparaiso University, Valparaiso, IN 46383}
\affil{todd.hillwig@valpo.edu}
\author{David J. Frew\altaffilmark{2,3}}
\altaffiltext{2}{Department of Physics, The University of Hong Kong, Pokfulam Road, Hong Kong SAR, China}
\altaffiltext{3}{Laboratory for Space Research, The University of Hong Kong, Pokfulam Road, Hong Kong SAR, China}
\author{Nicole Reindl\altaffilmark{4}}
\altaffiltext{4}{Department of Physics and Astronomy, University of Leicester, University Road, Leicester, LE1 7RH, United Kingdom.}
\author{Hannah Rotter\altaffilmark{1,5}}
\altaffiltext{5}{Current address: Department of Astronomy, San Diego State University, 5500 Campanile Drive, San Diego, CA 92182, USA}
\author{Andrew Webb\altaffilmark{1}}
\author{Steve Margheim\altaffilmark{6}}
\altaffiltext{6}{Gemini Observatory, Southern Operations Center, Casilla 603, La Serena, Chile}

\begin{abstract}

We find the central stars of the planetary nebulae (PNe) HaTr~7 and ESO~330-9 to be close binary systems.  Both have orbital periods of less than half a day and contain
an irradiated cool companion to the hot central star.  We provide light curves, spectra, radial velocity curves, orbital periods, and binary modeling results for both systems.  
The binary modeling leads to system parameters, or ranges of allowed parameters for each system.  We find that for the CS of HaTr~7 we need to use limb-darkening
values for the cool companion that are different than the expected values for an isolated star.  We also fit the central star spectrum to determine $\log g$ and
temperature values independent of the binary modeling.  For ESO~330-9 we find that based on our binary modeling the hot central star is most likely a post-RGB
star with a mass of around 0.4 M$_\odot$.  We discuss our derived stellar and nebular parameters in the broader context of
close binary central stars and their surrounding PNe.  
 We also discuss the present status of known or suspected post-RGB stars in PNe.

\end{abstract}

\section{INTRODUCTION}

As the sample of known close binary central stars of planetary nebulae (CSPNe) increases, we are able to begin to compare characteristics of the central stars (CSs) and
companions.  For example, with the exception of Hen~2-428 \citep{san15}, all existing models give CSs that are post-AGB stars consistent with existing stellar 
evolution models \citep[e.g.][]{sch83,blo95}.  However, there is some evidence that a handful of PNe may have post-RGB CSs (see \S 4 for further discussion).
We also find that cool companions in close binary CSPNe are typically over-luminous compared to their main sequence (MS) counterparts, presumably due to the irradiation effect from the hot nearby CS.  To date, no brown dwarf or planetary companions have been found in close orbits with CSPNe, though an increasing percentage of double degenerate (DD) systems are being discovered and it is possible that there are as many close binary CSPNe with white dwarf (WD) companions as those with MS companions \citep[c.f.][]{hil14}.

We can also compare resulting binary system parameters such as orbital period distribution and orbital inclination relative to the surrounding planetary nebula (PN).
The orbital period distribution of known close binary CSPNe is similar to systems without a PN (either because the ejected envelope was never visible or
because the system is past the PN stage), though it is possible that the CSPN systems have a slightly longer period distribution \citep{hil11}.  Both distributions
show very few systems with periods greater than a few days, suggesting that the common envelope (CE) phase is very efficient in reducing the orbital separation
of the two stars.  This is consistent with recent work on CE evolution \citep[e.g.][]{ric12,pas12,nan15,iac16,ohl16}.  Though broad examination of possible initial
binary parameters has not been performed due to complexities of the CE evolution process and limitations on computational abilities, significant headway is being made
on our understanding of the CE process.  With further work in both CE simulations and observational determination of stellar and orbital parameters of close binary CPSNe,
more detailed and statistically significant comparisons will be possible.

\citet{hil16b} have shown a clear link between the inclinations of close binary CSPNe and their associated PNe, demonstrating a physical connection between the two.
Such a connection has long been suspected (and expected in some cases) with a number of papers showing correlations between, e.g. strongly bipolar PNe and
post-CE CSs \citep{zij07, dem09, mis09b}.

Apart from studies of the stars themselves, work is being done on relating the nebula around close binary CSPNe.  \citet{cor15} show a relationship
between close binary CSPNe and abundance discrepancies in their associated PNe \citep[see also][]{jon16}.  Considering the abundance discrepancies
and low ionized mass of many of these PNe
\citep[also discussed by, e.g.][]{fre08,fre07} they suggest the possibility that these may not be PNe at all, but some later stage of mass loss or even the disruption of a
Jupiter-mass planet.

One goal of this series of papers \citep*{dem08} has been to use photometric monitoring to discover additional close binary CSPNe to improve the binary parameter statistics.
Here we present two newly discovered systems: the CSs of HaTr~7 and ESO~330-9.  Both are irradiated binary systems with cool
companions and have orbital periods of less than half a day.  Below we show multi-color photometry, spectra, and radial velocity curves for both systems and calculate
parameter sets for both using the Wilson-Devinney binary modeling code \citep{wil71,wil90}.

\section {The Central Star of the Planetary Nebula HaTr 7}

\subsection {Background}

The faint PN Hartl-Tritton 7 (HaTr~7, PN G332.5-16.9) was discovered on UK Schmidt plates by \citet{har85}.
They describe the nebula itself as a ``slightly elliptical disk with very low surface brightness''.

\citet{sau97} provide a spectrum of the CS of HaTr~7 showing both absorption from the CS and emission lines
near 4650 \AA.  This type of emission in CSPNe has often lead to a ``weak emission line star'' (wells) classification.
More recently it has become understood that the wels classification is not a separate physical effect \citep[e.g.,][]{wei15}, but a combination
of unrelated causes such as irradiation effects \citep[e.g.,][]{mis11}, residuals from nebular subtraction \citep[e.g.,][]{bas16}, and other effects. 
\citet{sau97} suggest these lines may be due to carbon enrichment, but we show below that they are irradiation lines from the
heated hemisphere of a cool companion.  They use a non-LTE model atmosphere including H and He to determine a temperature
for the CS of $T_{CS}=100,000$ K, a surface gravity $\log g=6.0$, and a CS mass of $M_{CS}=0.56$ M$_\odot$.

\begin{figure}[p]
\begin{center}
\includegraphics[width=6in,angle=-90]{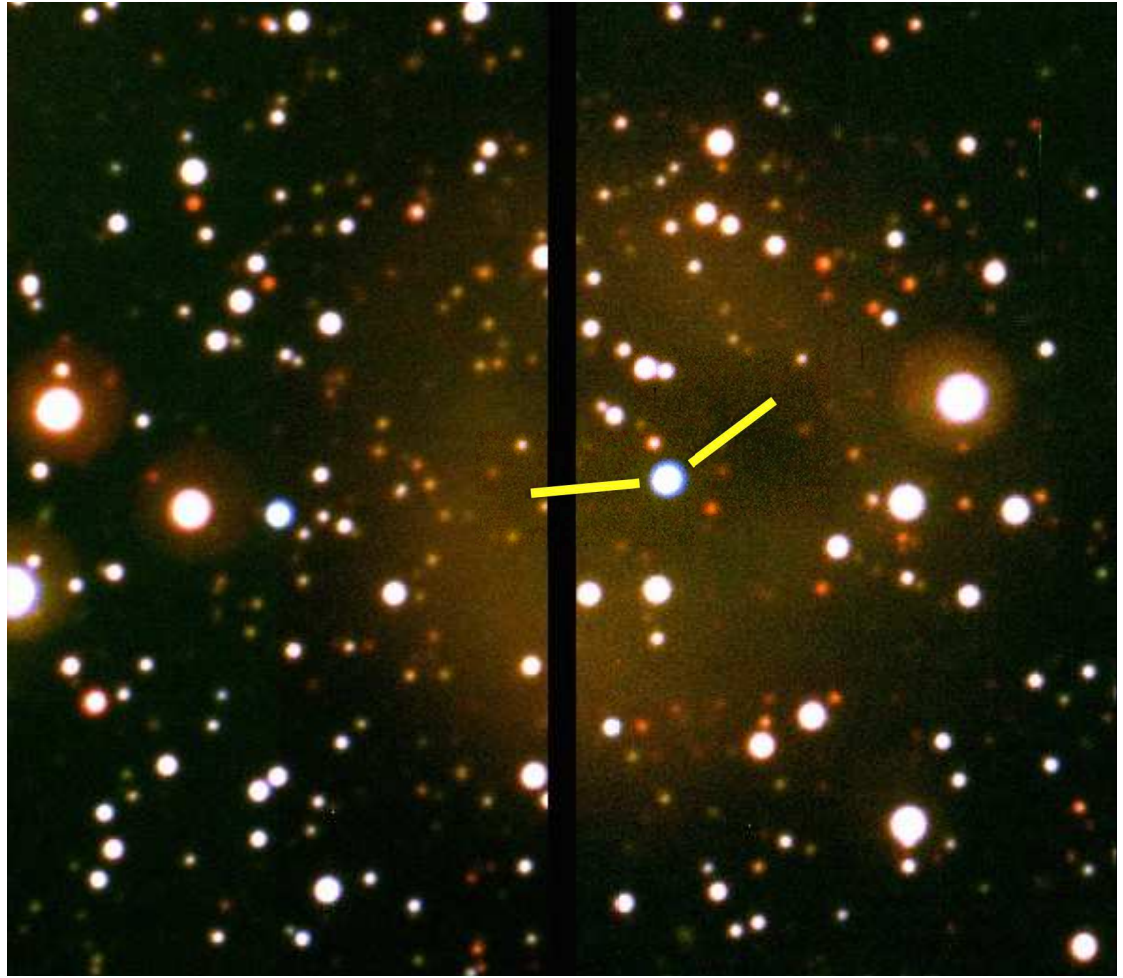}
\end{center}
\caption[HaTr 7 Color Image] {A color composite image 
of HaTr 7 from the SOAR telescope.  The CS is the bright blue star at the geometric center of the PN identified by hash marks in the image.  
North is up and East is to the left.
\label{hatr7img}}
\end{figure}

We provide a color composite image of the nebula in Figure \ref{hatr7img}.  The images were obtained with
the SOAR 4-meter telescope.  Red is from an H$\alpha$ image, green from an \ion{O}{3} image, and blue from a $B$ filter image.
The dark stripe in the center of the image is from the inter-chip gap on the CCD.  While the nebulosity is faint, the general shape
looks similar to that of Abell~65 \citep{hil15} which \citet{huc13} found to be a double-lobed PN at an angle of about 62$^\circ$.  It is possible that
HaTr~7 has a similar shape and is also at an intermediate angle.

\citet{chu09} provide a {\it Spitzer} MIPS 24 $\mu$m band image of HaTr~7.  Their image looks very similar to our visible image though with
a more pronounced southern ``lobe'' and a smaller spatial extent.  They do not find a point-source at the location of the CS, suggesting that it does not have a strong
dust component to its spectrum.

\citet{fre16} give a distance to HaTr~7 of $1.85\pm0.53$ kpc and an interstellar reddening value of $E(B-V)=0.08\pm0.03$.

Below we explore in more detail potential binary models and compare our results to the information discussed above. 

\subsection{Observations and Reductions}

The photometric data consist of $B$, $V$, and $R$ images obtained with the SARA-CTIO 0.6 meter telescope in 2014 April, 2014 May, and 2015 June.  
We also utilize 82 $V$ data points from the Catalina Sky Survey \citep[CSS,][]{dra09}.

All images were bias subtracted and flat fielded.  The IRAF/DAOPHOT\footnote{IRAF is distributed by the National Optical Astronomy Observatory, which is operated by the Association of Universities for Research in Astronomy (AURA) under a cooperative agreement with the National Science Foundation.} package was used to perform aperture photometry on the images.  Single star differential photometry was performed on the resulting data with two additional stars used to check the photometric stability of the comparison star.  We find that the comparison star is constant with standard deviations from the mean of 0.003 mag in $B$, 0.005 mag in $V$, and 0.004 mag in $R$.

\begin{figure}[p]
\begin{center}
\includegraphics[width=6in]{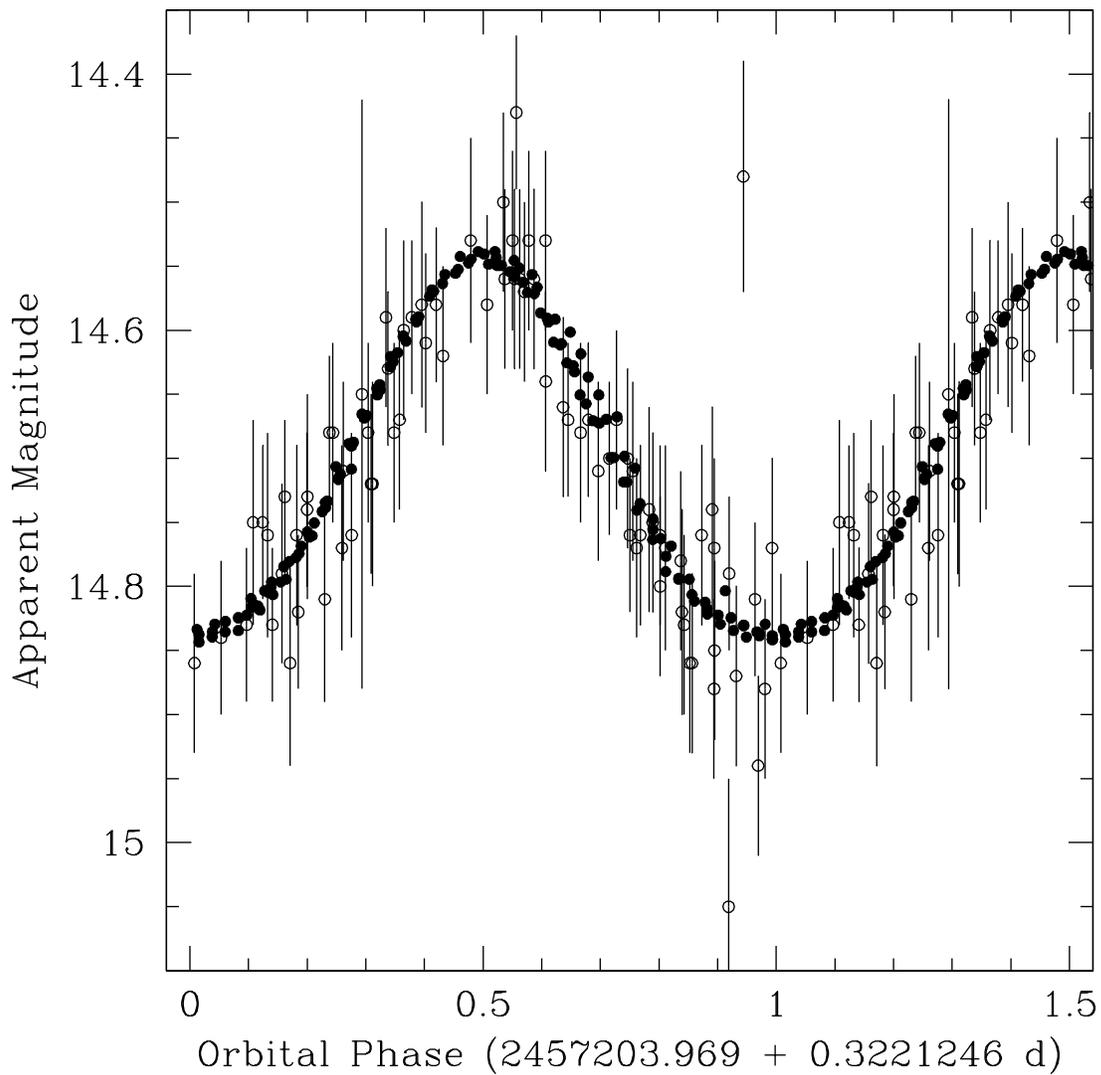}
\end{center}
\caption[HaTr 7 Phase-Folded CSS and SARA V Light Curve] {The $V$-band photometry from CSS ({\it open circles}) and SARA-CTIO ({\it filled circles})
phase-folded onto the ephemeris in the text.  The SARA-CTIO data has been shifted in magnitude to match the CSS apparent magnitude scale.
\label{hatr7css}}
\end{figure}

\begin{figure}[p]
\begin{center}
\includegraphics[width=6in]{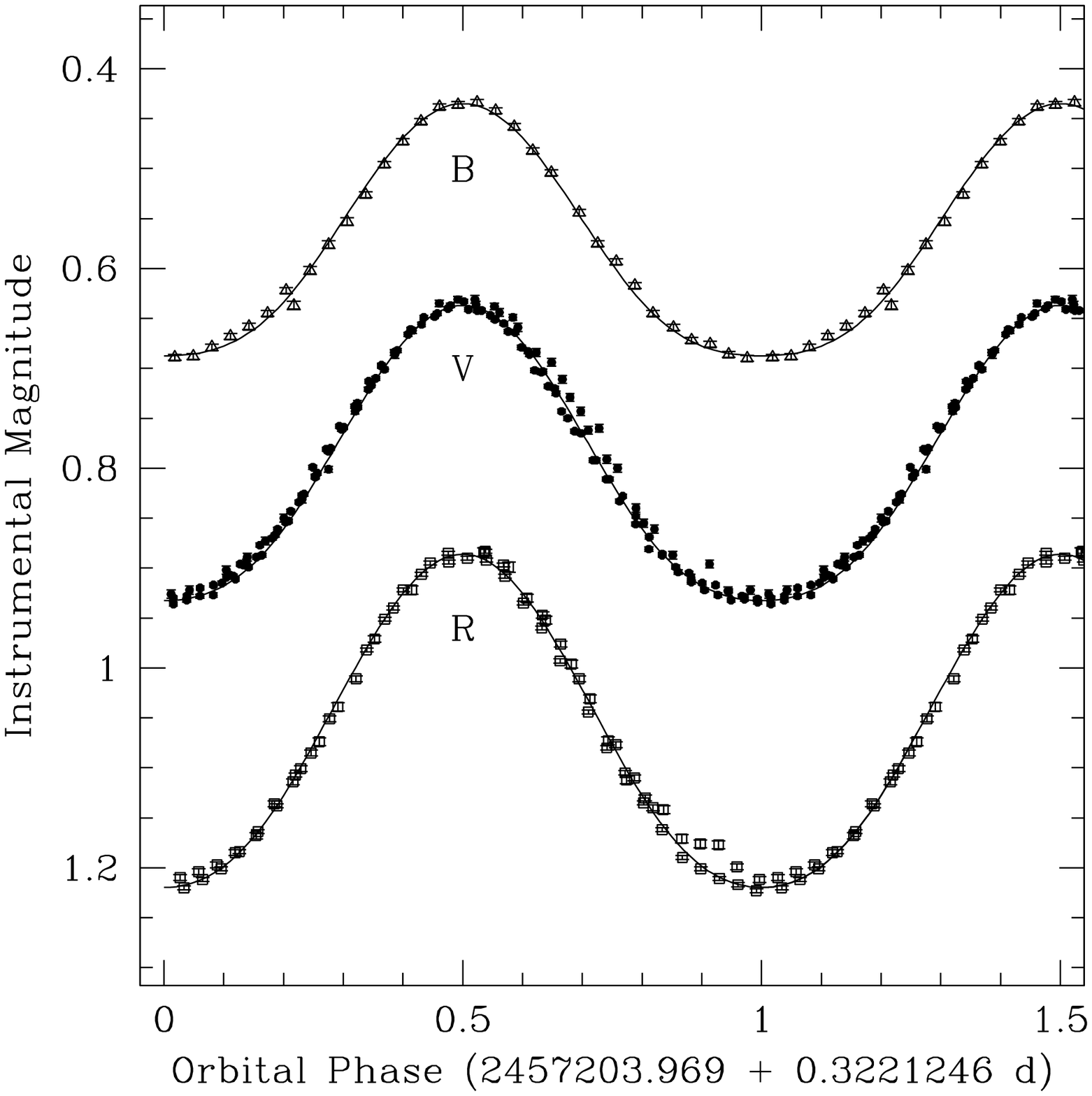}
\end{center}
\caption[HaTr 7 Phase-Folded Light Curve] {The differential magnitude $B$, $V$, and $R$ phase-folded light curves
of HaTr 7 for the ephemeris in the text.
The solid lines correspond to a binary model from the Wilson-Devinney code as described
in the text.  Error bars are included though in most cases they are smaller than the symbols.
\label{hatr7phot}}
\end{figure}

Using the long baseline of our photometric data and the CSS data together gives a photometric ephemeris of
$$T=2457203.970(5) + 0.3221246(8)\times E \textrm{ days}.$$
Here $E$ is the number of orbits since time $T0$, with $E=0$ corresponding to minimum light measured in heliocentric Julian date.
In Figure \ref{hatr7css} we show the CSS $V$ data along with our $V$ data, where we have shifted our data to the apparent magnitude scale of the CSS data.
Figure \ref{hatr7phot} shows the
$B$, $V$, and $R$ light curves for HaTr~7 along with an example binary model (described further below).  We do not include the CSS $V$ data in this plot
because of the larger scatter and uncertainties and for visibility of the model lines compared to the data.

We also obtained orbit-resolved spectra in 2016 June using GMOS on the Gemini-South telescope in longslit mode with the B1200 grating and a 0.75$\arcsec$ wide slit.  
The spectra cover a range of 3825-5442 \AA~ with a
resolution of $R=3744$ and 2x2 binning resulting in 0.516 \AA~ per pixel.  The IRAF/Gemini package was used to reduce the spectra.
Wavelength calibration used CuAr arc spectra taken consecutively
with the science spectra, giving a typical radial velocity calibration of approximately 0.2 \AA~or better.

\begin{figure}[p]
\begin{center}
\includegraphics[width=4.5in,angle=-90]{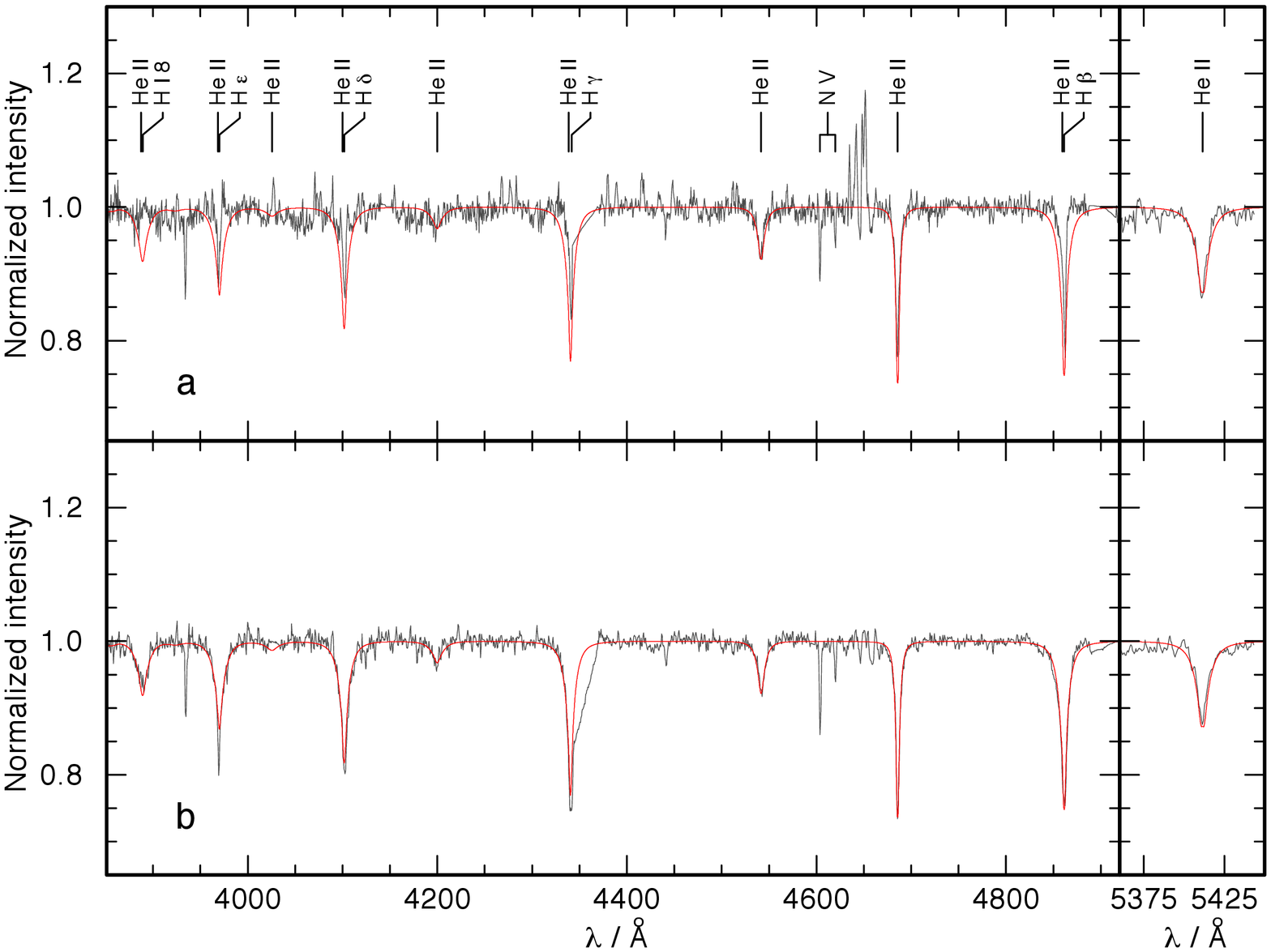}
\end{center}
\caption[HaTr 7 Spectrum] {Two sample spectra of the CS of HaTr 7 taken near maximum light (a) and minimum light (b).  Easily visible is
the CNO complex of emission lines near 4650 \AA~due to irradiation
of a cool companion.  Absorption lines from the hot CS are also present and dominate the spectrum in (b).  We also show our TMAW
model fits giving with $T=95300\pm2000$ K and $\log g=5.79\pm0.04$ in red.  Chip gaps near the H$\gamma$ and H$\beta$ absorption features made continuum
fitting near these lines difficult in the red wings of the lines.
\label{hatr7spec}}
\end{figure}

A spectrum near maximum light (Figure \ref{hatr7spec}a) shows strong emission lines consistent with an irradiated companion.  In addition, there are 
\ion{N}{5} $\lambda\lambda$ 4604, 4620 \AA, \ion{He}{2}, and hydrogen Balmer absorption lines from the hot CS.  The spectrum in Figure \ref{hatr7spec}b obtained
close to minimum light shows a near absence of emission lines, suggesting an irradiated binary at a moderate to high inclination such 
that the hottest portion of the cool star, the sub-stellar point nearest the hot CS, moves out of view for that portion of the orbit.  
The minimum light spectrum also provides an opportunity to reassess the 
model atmosphere of \citet{sau97}.  Given the strong emission lines near 4650 \AA~in their published spectrum, it must have been obtained near maximum light.
For our observed photometric amplitude, at that orbital phase the heated secondary is contributing as much as 30\% of the system light.  Thus the absorption lines will be
weakened by the additional light source.  If we assume that the light from the secondary is negligible at minimum light, the line depths are then representative of the
CS alone.  

For the quantitative spectral analysis we used the model grid from \citet{rei16}, which was calculated with the T\"ubingen non-LTE model-atmosphere
package  \citep[TMAP\footnote{http://astro.uni-tuebingen.de/ TMAP},][]{wer03, rau03}. The parameter fit was performed by means of a $\chi^2$ 
minimization technique with SPAS (Spectrum Plotting and Analysing Suite, \citealt{hir09}), which is based on the FITSB2 routine \citep{nap99}. 
We fitted all Balmer and \ion{He}{2} lines detected in the spectrum of HaTr 7. Because of chip gaps overlapping the red wing of H$\gamma$ and falling very near 
the red wing of H$\beta$, we fit only the blue-ward half of those two lines. Our results give $T=95300\pm2000$ K, $\log g=5.79\pm0.04$, and a log Helium mass 
fraction $X_{He}=-0.56\pm0.03$, close to the solar value. These are in good agreement with \citet{sau97}. In Figure \ref{hatr7spec} we compare our best fit model 
with the observations. Comparing the TMAP model with the spectrum taken at maximum light (\ref{hatr7spec}a), the impact of the light of the companion becomes obvious.
While the pure \ion{He}{2} line profiles are hardly affected, the the Balmer lines become much more narrow and their intensity is reduced by about 15\%. 

\begin{figure}[p]
\begin{center}
\includegraphics[width=6in]{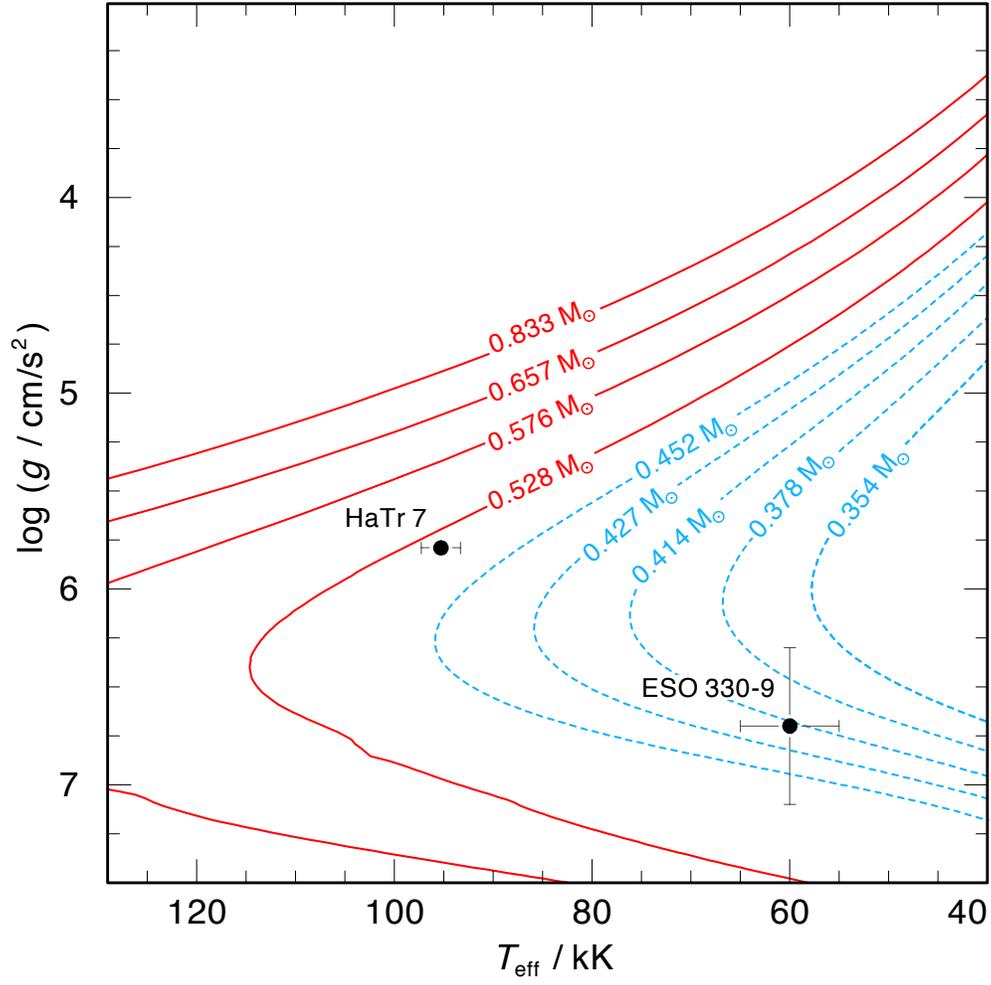}
\end{center}
\caption[CSs of HaTr~7 and ESO~330-9 in log(g) vs. T] {Our model results of $\log g$ and $T$ for the CSs of HaTr~7 and ESO~330-9 plotted
with the post-AGB evolutionary tracks of \citet{mil16} ({\it red solid lines}) and post-RGB tracks of \citet{hal13} ({\it blue dashed lines}).
\label{postAGBRGB}}
\end{figure}

To derive the mass and luminosity of the stars, we compared the position of HaTr 7 in the $T_{eff}$ -- $\log g$ plane with latest H-burning post AGB tracks of 
\citet{mil16} (Figure \ref{postAGBRGB}). We find that the central star should have a mass of 0.52 M$_\odot$ and $L=2000$ L$_\odot$, i.e. 
$\log (L/L_\odot)= 3.3$.   

\begin{figure}[p]
\begin{center}
\includegraphics[width=6in]{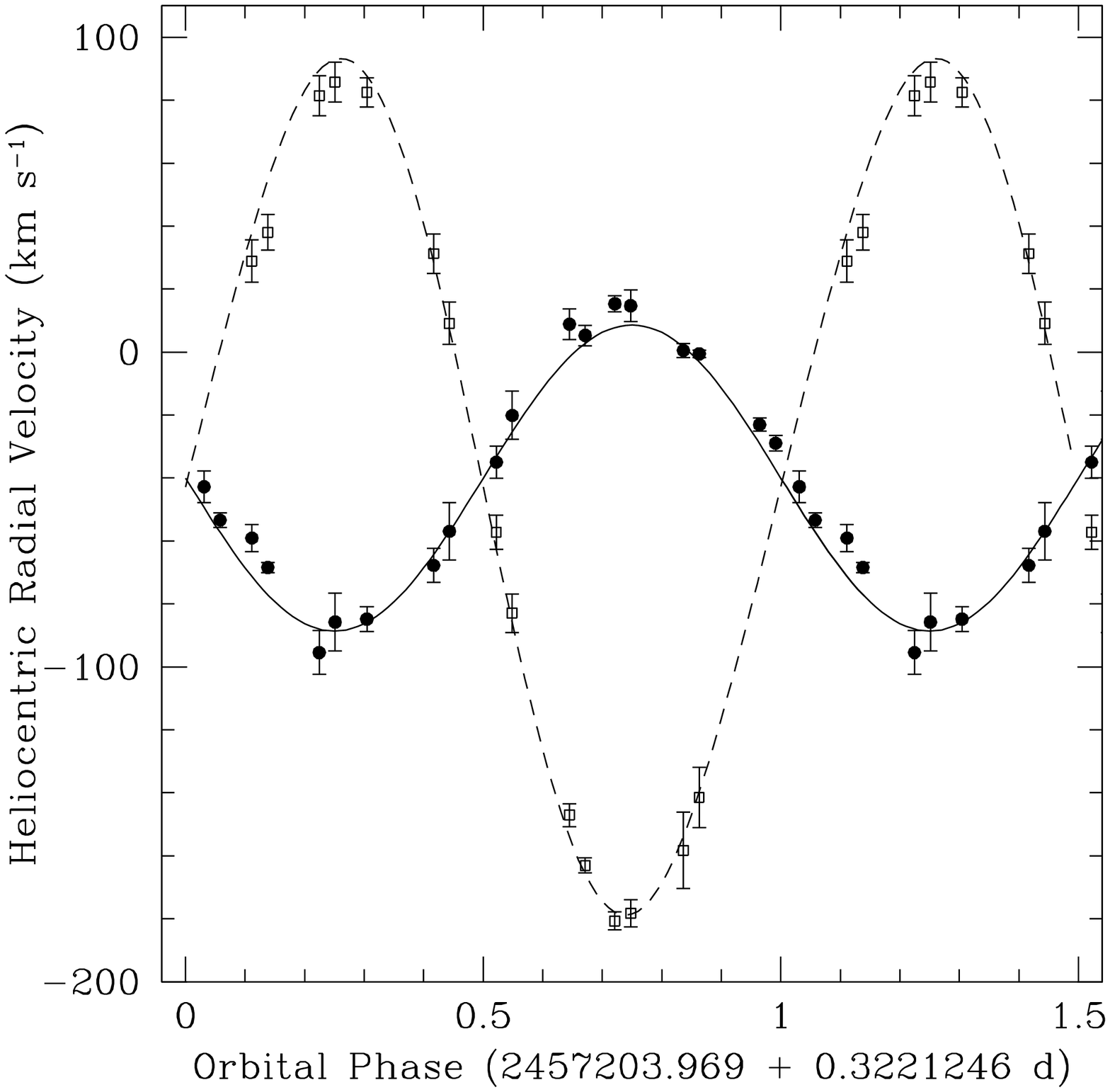}
\end{center}
\caption[HaTr 7 Phase-Folded Radial Velocity Curve] {The radial velocity curve for the CS of HaTr 7.  Data points are for the irradiation lines ({\it hollow squares})
and absorption lines ({\it solid circles})
described in the text.  The dashed line is the radial velocity curve for the irradiated companion, the solid line is for the CS, based on our Wilson-Devinney
modeling.
\label{hatr7rv}}
\end{figure}

To determine radial velocities for the CS and companion we fit Gaussian profiles to the narrow emission lines and Voigt profiles to the broad absorption features.
The C, O, and N emission lines near 4650 \AA~are the only strong irradiation lines in the spectrum, so we used the average velocities for those lines to determine a
radial velocity for the companion for each spectrum.  For the CS we used the average velocity of the visible absorption lines of \ion{H}{1}, \ion{He}{2}, and \ion{N}{5}.
The resulting radial velocity curve (Figure \ref{hatr7rv}) confirms that the photometric period is equal to the
orbital period.  We also see that the emission lines do originate from the companion star, confirming their origin in an irradiated hemisphere of a cool companion.

Sine curve fitting to the CS and companion curves result in velocity zero-points that are different by about 11 km s$^{-1}$.  We address this issue
in our binary modeling section below.

\subsection{Binarity and System Parameters}

For our binary modeling we adopt initial values for the CS mass, $\log g$, and temperature values in the range of
our spectral fitting and that of \citet{sau97}.  The primary difference comes in the CS mass as described above.
So we use $0.52\leq M_{CS}\leq 0.56$ M$_\odot$, $5.5\leq \log g\leq0.60$, and $90,000\leq T_{CS}\leq 100,000$ K.
The double-lined radial
velocity curve allows us to determine a mass ratio, however we need to be careful about simply using the ratio of the
amplitudes of the two curves.  Because the companion radial velocity curve is determined using emission lines due to
irradiation, the amplitude traces the velocities of the irradiated hemisphere and not the stellar center-of-mass.  The amplitude then
depends on the radius of the secondary relative to the star's distance to the system center-of-mass.  Therefore, rather than
adopting a center-of-mass velocity, we require that the resulting model produce a good fit to the radial velocity data.  The ratio of the
two amplitudes from sine curve fitting is 0.38, which we use as a maximum expected value for the mass ratio, $q=M_2/M_{CS}$
(assuming that the center-of-mass amplitude of the companion is larger than that found from the emission lines).

As mentioned above, we find that fitting independent sine curves to the two radial velocity curves produces a zero-point
offset of about 11 km s$^{-1}$ with the CS at a more redshifted velocity than the companion.  Some of this will be due to
gravitational reddening from the CS, which in turn depends on the
mass and radius of the CS.  However, using the mass and $\log g$ values from \citet{sau97} and our modeled values gives a 
resulting gravitational redshift in the range 2--3 km s$^{-1}$, which only corrects
about one-quarter of the observed shift.  The cause of the remaining shift is unclear.  Zero-point offsets in the spectra should produce
identical behavior in the CS and companion, rather than opposite shifts.  It is possible that the emission lines from the
companion are being created high in the atmosphere and that a wind could produce an additional blue-shift.  These are
speculation however and without knowing if it is an intrinsic or data cause, we correct for the CS gravitational redshift and
proceed with the binary modeling to achieve the best possible fit to the existing data.

The first result we find from our binary modeling is that for the temperature of the CS and relative sizes of the CS and companion,
the temperature of the companion is not well constrained.  Any temperature below about 5000 K produces almost identical
light curve amplitudes.  We believe that this effect is essentially due to a ``saturation'' of the irradiation effect in the atmosphere
of the companion.  That maximum temperature does change slightly depending on the rest of the system parameters, but
we find that a companion temperature of 4500 K is indistinguishable from lower temperatures for all of our remaining parameter sets.
Thus for all of our models we set the companion temperature at 4500 K.

From Figure \ref{hatr7phot} we see that the light curve amplitudes in $B$, $V$, and $R$ are measurably different, with
the $R$ amplitude being the largest and $B$ the smallest.  For these types of binaries, this trend in amplitudes usually
means that the companion provides a significant contribution to the total light of the system even disregarding the
irradiation effect.  However, the small orbital period and small mass ratio give a correspondingly small orbital separation
and Roche lobe radius for the companion, meaning that the maximum volume radius for the companion should be less
than 0.5 R$_\odot$.  In this case, the companion would {\it not} contribute significant light to the system.

Proceeding with our binary modeling we indeed find that for realistic companion stars we cannot reproduce the relative
light curve amplitudes by adjusting only the inclination and stellar masses, radii, and temperatures.  We find that
for realistic stellar parameters, when the binary model light curves fit the observed $B$ amplitude, the $V$ and $R$
amplitudes are too small (the $R$ being more discrepant than the $V$).  And vice versa, when we match the
model curves to the $R$ filter, the $V$ and $B$ model curves have amplitudes larger than the data.
Only for CS temperatures {\it much} lower than those consistent with the CS spectrum and its
lack of \ion{He}{1} lines (requiring $T_{CS}>$70,000 K) do we find
matches for the relative amplitudes.  The only means of producing a parameter set that matched the data was to
use limb darkening coefficients different than those calculated internally by the Wilson-Devinney code.  While there have been
close binary CSPNe for which different limb-darkening coefficients have been reported, even limb-brightening \citep[e.g.,][]{bon87,pol94}, as a default we use
the internally calculated values unless, in situations like this, it is clear that those values will not provide a match to the data.
The difficulty with changing limb darkening coefficients is that at this point those changes are purely empirical.  Without a better
understanding of the physical changes in limb darkening any changes are essentially arbitrary and can produce
very large changes in the resulting light curve amplitudes.  We choose to modify limb-darkening values rather than
another stellar parameter or parameters such as albedo or gravity darkening because, as described above, previous studies have 
suggested that limb-darkening of the cool companions in these systems may be different than the expected values for their MS equivalents.
And it is easy to understand physically how limb-darkening could be different for an irradiated atmosphere in which, if the effect is large enough,
a temperature inversion may take place in the stellar photosphere.

We initially used a quadratic form for the limb darkening, but find that a linear form
produces nearly identical results and provides a better intuitive understanding of the required changes.
Here we use the linear form $I(\mu)/I(1) = 1-u(1-\mu)$ from \citet{cla00} where $I(1)$ is the intensity at the center of
the stellar disk, $\mu=\cos\gamma$, and $\gamma$ is the angle from the line of sight from the stellar center to the point of 
emission on the stellar surface.  Thus smaller values of $u$ result in a less limb darkened disk, values of $u>1$ are not
physical, and values of $u<0$ would result in limb brightening.
In this case, we find that the more red filters require larger negative changes in limb darkening than the bluer filters.
Thus for example, if we leave the $B$ filter limb darkening at the expected value for the companion star, the $V$ and $R$ limb
darkening values must be reduced, such that the $V$ limb darkening is somewhat reduced from its expected value and the
$R$ limb darkening is significantly below its expected value.  In other words, the redder the wavelength, the less
the star is limb-darkened.  Using the linear limb-darkening coefficients from \citet{cla00} we find values for the companion
of $u_B=0.881$, $u_V=0.796$, and $u_R=0.723$ for a temperature of 4500 K and $\log g = 4.5$.  In order to fit the
data, keeping $u_B$ constant, we find values of $u_V=0.610$ and $u_R=0.152$.  Allowing $u_B$ to change
as well results in different modeled radii for the companion.  Reducing the limb darkening further results in smaller radii, while
increasing $u_B$ results in larger radii.  The radius changes by less than 10\% in each direction unless the limb-darkening
values become negative as they decrease, or unphysical as they increase.

We provide in Table \ref{hatr7model} a range of system parameters which provide acceptable fits to the data.  
We find that the CS agrees with evolutionary models of post-AGB stars, as expected based on using values consistent with
model spectra.  Our modeled parameter range also includes our spectroscopically determined values for the CS in
temperature, radius, $\log g$, and luminosity.  In Figure \ref{postAGBRGB} we plot the CS in $\log g$ vs temperature along
with post-AGB models of \citet{mil16} and post-RGB models of \citet{hal13}.
Each of the parameter combinations we found results in a companion star larger than we would
expect for a MS star of its mass.  Given our inability to determine a precise temperature for the secondary, we cannot speak directly to
how its temperature would compare to that of a similar MS star.  However, the upper limit to the temperature is higher
than equivalent MS temperatures for all ranges of our companion star parameters.  So our companion temperature is consistent
with or slightly higher than for an equivalent MS star.
Over-luminous companions due to larger radii and/or higher temperatures are typical in the cool companions of close binary CSPNe.

The inclination of the binary system we find to be $i = 45$ -- 50$^\circ$.  While we do not have an inclination for the PN, our visual
inspection described above did suggest an intermediate inclination.  Thus our binary inclination is at least consistent with that expectation.

Using the $E(B-V)$ value from \citet{fre16} and a measured apparent magnitude we can use our model to calculate a distance
to HaTr~7.  The available apparent visual magnitudes are consistent within our observed 0.3 magnitude amplitude variability in the light curve.
The SPM~4 catalog \citep{gir11} gives an estimated $V=15.03$, the UCAC~4 catalog \citep{zac12} gives $V=15.111$, and the APASS value is
$V=14.954\pm0.232$ \citep{hen16}.  However, we do not know at what phase these data were obtained.
Using the data from the Catalina Sky Survey, we find a minimum-light value of $V=14.84\pm0.08$ where the error is based on both the
scatter in the light curve and the zero-point calibration of the CSS data.  While this minimum value is slightly brighter than the SPM~4, UCAC~4, and
APASS values, they are consistent given the uncertainties of those measurements.  
Using the minimum-light value from the CSS data of $V=14.84$ and using $E(B-V)=0.08$, we find
a de-reddened apparent visual magnitude of the CS of HaTr~7 of $V=14.57\pm0.08$.
From our binary model we find a total system brightness at minimum light of $2.5<M_V<3.2$.  From these values we calculate a distance
range from our model of $1.9<d<2.6$ kpc, which agrees well with the distance of $1.85\pm0.53$ kpc from \citet{fre16}.
\citet{fre16} also used the atmospheric parameters from \citet{sau97} to determine a spectroscopic ``gravity?? distance of 
$1.80 \pm 0.70$ kpc (see their Table~5).  Using our revised parameters for the ionizing star, a nearly identical distance of $1.90\pm0.50$ kpc 
is calculated, our lower gravity estimate essentially canceling out the brighter $V$-band magnitude we now use.  This value is also in  very good 
agreement with the surface brightness distance.

\begin{center}
\begin{deluxetable}{lc}
\tablewidth{0pc}
\tablecolumns{3}
\tablecaption{HaTr 7 Binary System Parameter Ranges
\label{hatr7model}}
\tablehead{
\colhead {Parameter}        & \colhead{Value}}
\startdata
$q$					& 0.30 -- 0.35 \\
$M_{CS}$ (M$_\odot$)	& 0.50 -- 0.56 \\
$M_2$ (M$_\odot$)		& 0.14 -- 0.20 \\
$T_{CS}$ (K) 			& 90,000 -- 100,000  \\
$T_2$ (K) 			& $\lesssim 5000$	\\
$R_{CS}$ (R$_\odot$)	& 0.125 -- 0.180 \\
$R_2$ (R$_\odot$)		& 0.3 -- 0.4 \\
$i$ ($^\circ$) 			& 45 -- 50 \\
$a$ (R$_\odot$)		&1.73 -- 1.79 \\
\enddata
\end{deluxetable}
\end{center}

\section {The Central Star of the Planetary Nebula ESO~330-9}

\subsection {Background}

\begin{figure}[p]
\begin{center}
\includegraphics[width=6in,angle=-90]{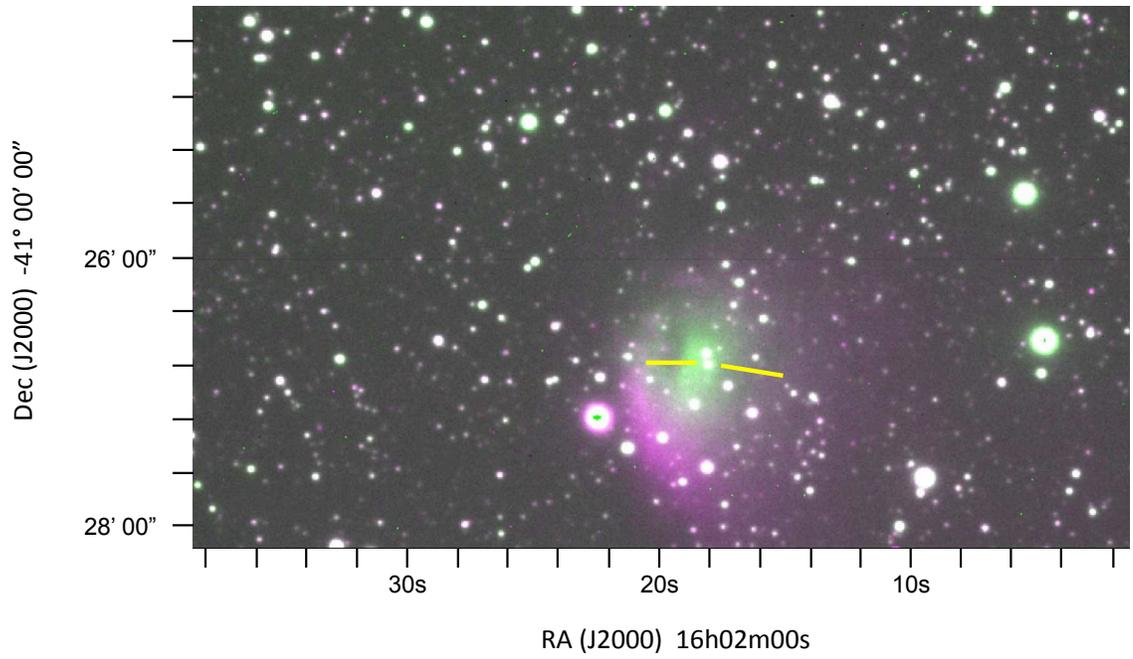}
\end{center}
\caption[ESO~330-9 Color Image] {A tricolor RGB image 
of ESO~330-9 from the POPIPLAN survey [\ion{O}{3}] ({\it green}) and H$\alpha$+[\ion{N}{2}] ({\it red and blue}) images.  The CS is identified by hash
marks in the image. \label{ESO330-9img}}
\end{figure}

ESO\,330-9 (PHR~J1602-4127; PN~G331.0+08.4) was first noted on the ESO Sky Survey \citep{lau82}, and described as either an emission 
nebula or galaxy.  It was re-discovered and identified as a planetary nebula in the MASH Survey by \citet{par06}, though \citet{mag03} 
catalogued it as a reflection nebula (GN~15.58.9).  MASH spectroscopy reveals the nebula to have a relatively modest excitation \citep[see][]{fre08} 
and narrowband images are additionally presented by \citet{bof12}.  ESO~330-9 is currently classified as a likely PN by \citet{par16}.
\citet{fre08} also gave a distance of 1.65 kpc 
and noted that based on available data the CS appeared to
have an IR excess and was thus a good candidate for having a cool companion.
Very little further work has been
done on the PN or central star.  Figure \ref{ESO330-9img} shows a color image of the nebula from the POPIPLAN survey \citep{bof12}
using [\ion{O}{3}] and H$\alpha$+[\ion{N}{2}] images from the VLT.  The CS, which was
identified by MASH is off-center relative to H$\alpha$+[\ion{N}{2}] (purple in the image), but is relatively well-centered in \ion{O}{3} (green).  
The abrupt and brightened south-eastern edge suggests interaction with the ISM, which may be the cause of the asymmetry.  A dark
lane is also visible near the edge of the central [\ion{O}{3}] emission.

\citet{fre16} give a mean distance of $1.73\pm0.51$ kpc to ESO~330-9 and an $E(B-V) = 0.27\pm0.10$.  However, the line-of-sight reddening from
\citet{sch11} is $E(B-V) = 0.64$.  We explore this further below. 

The reported visual apparent magnitude of the CS 
from the SPM~4 catalog \citep{gir11} is $V=16.86\pm0.2$.  However, the YB6 catalog\footnote{YB6 = Yello-Blue all-sky catalog, version 6: 
Monet, D.G. 2004, complete scan of USNO NPM and SPM plates} gives $V=15.76\pm0.4$ so it is difficult to determine from these a
precise value for the visual magnitude.  DENIS and 2MASS give more reliable values for the near-IR magnitudes $I = 16.04 \pm 0.06$ (DENIS),
$J = 15.07 \pm 0.14$ (DENIS), $J = 15.36 \pm 0.05$ (2MASS),
$H = 14.54 \pm 0.07$ (2MASS), and $K_s = 14.38 \pm 0.07$ (2MASS).  The difference in the DENIS and 2MASS $J$ magnitudes is not due to
intrinsic variability, as the orbital phase of the two measurements are only different by about 0.08.  The dates at which the measurements were
obtained are too far separated in epoch from our data to allow proper determination of the actual orbital phase, but even for phases providing the
maximum magnitude difference we find from our models over a 0.08 orbital phase difference $\Delta J=0.06$.  At that level
the two values marginally agree with one another.  We also find a GALEX measurement of the near-UV giving $m_{NUV}=17.97\pm0.045$.

There is no age determination for the PN.  Part of the difficulty in an age determination is the apparent
ISM interaction, which makes determining an age based on the size and expansion rate of the nebula very difficult.  There is also no published value for the
expansion rate of the nebula.  But the nebular size reported by \citet{fre16} of 200\arcsec~by
175\arcsec~gives an average radius of about 95\arcsec.  If we then assume a typical expansion speed of about 10 km s$^{-1}$, we find a
very approximate age of 78,000 years for the distance of 1.73 kpc.

\subsection{Observations and Reductions}

Our photometric data were obtained with the 0.6m SARA-CTIO telescope in 2013 April, May, and June in the $V$ filter and 2013 June and July in the $B$ and 
$R$ filters.
The images were bias subtracted and flat fielded.  The IRAF/DAOPHOT package was used to perform aperture photometry on the images then single-star
differential photometry was used to obtain a light curve.  Two additional check stars were used to verify the stability of the comparison star, which was constant to
0.015 mag in both $B$ and $V$ and to 0.008 mag in $R$.
We find an orbital ephemeris (assuming a binary system) of
$$T=2455294.6925(5)+0.29592(4) \times E \textrm{ days}.$$
where $E=0$ occurs at minimum light.

\begin{figure}[p]
\begin{center}
\includegraphics[width=6in]{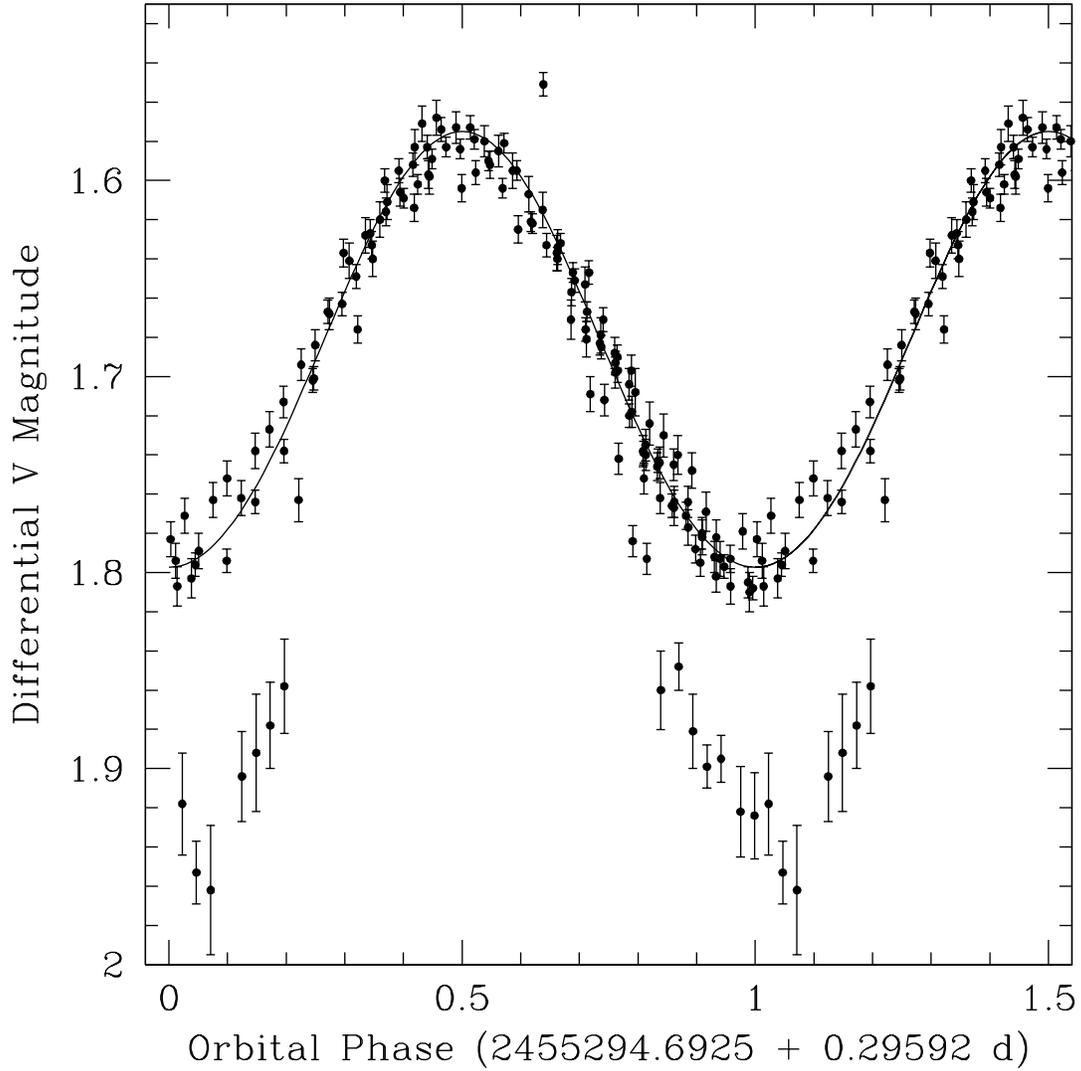}
\end{center}
\caption[ESO~330-9 Phase-Folded V Light Curve] {Phased differential $V$ magnitude light curve
of ESO~330-9 for the ephemeris in the text.  The one-time observed decrease in brightness is apparent near
minimum light.  The line is from the Wilson-Devinney
model fit as described in the text.
\label{ESO330-9lcV}}
\end{figure}

Figure \ref{ESO330-9lcV} shows our $V$-band light curve folded on the ephemeris above.  Our first set of observations at minimum light are significantly fainter
than the remaining observations at minimum light.  We initially thought that the system may show ellipsoidal variability with two different minima.  But later
photometry and radial velocity data confirmed the irradiation effect and shorter orbital period (equal to the photometric period rather than twice the photometric period).
We explored data processing sources for the fainter magnitudes in that portion of the one night of data, but ruled out throughput, seeing, or comparison star causes.
At this point we assume that the cause is intrinsic to the system, perhaps an orbiting dust cloud as seen in \citet{gro93} and \citet{haj08}, though the timescale
for the variability in those systems is much longer.  In this case the dust would have to be on a much smaller orbit, or the dust is being produced in a wind and is
not following a closed orbit.

\begin{figure}[p]
\begin{center}
\includegraphics[width=6in]{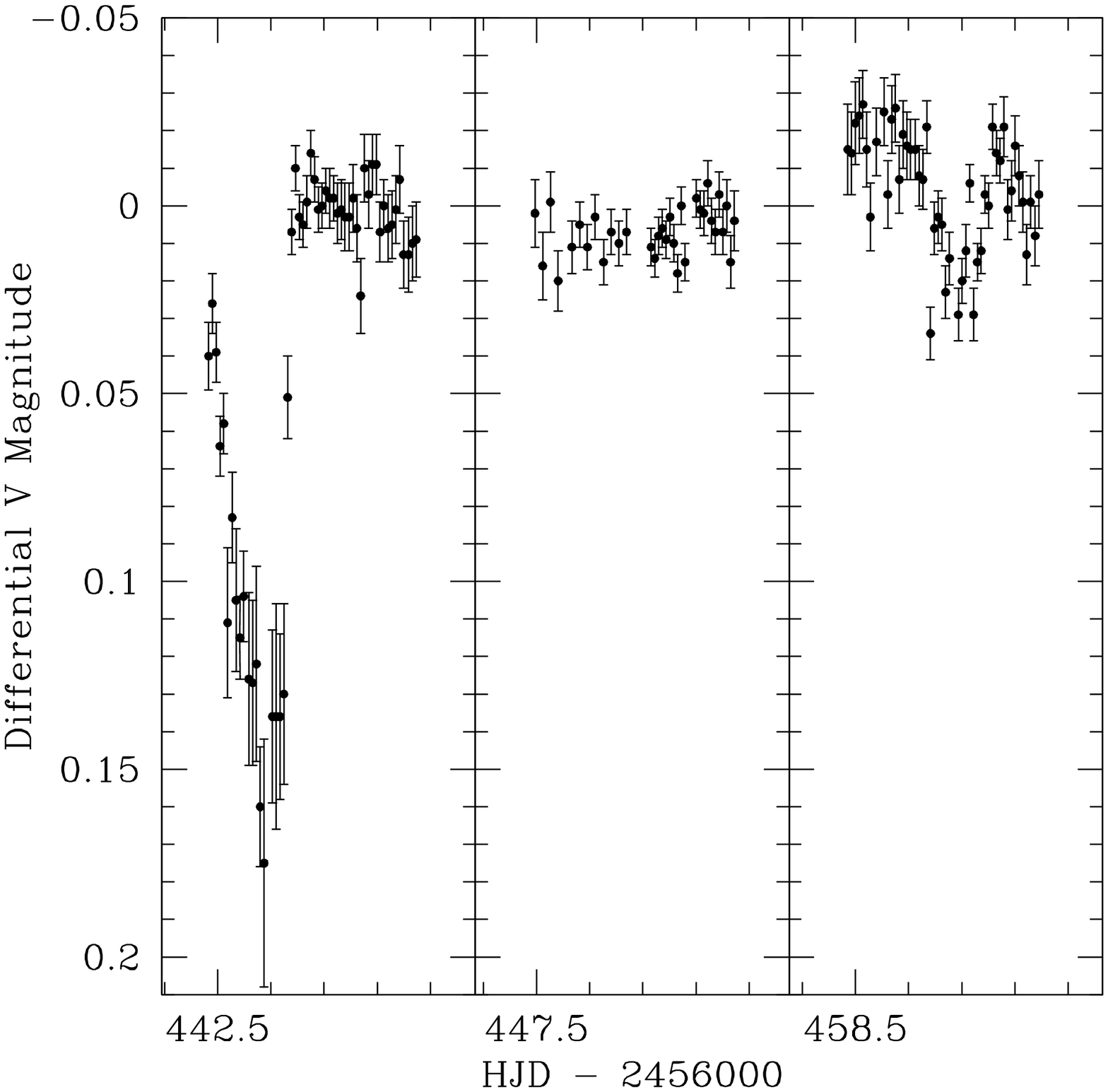}
\end{center}
\caption[ESO~330-9 Model Subtracted Light Curve] {Three nights of $V$-band data with our binary model curve subtracted to show
variability other than due to the close binary.  The small tick-marks on the x-axis denote steps of 0.1 day.
\label{ESO330-9Vdiff}}
\end{figure}

In Figure \ref{ESO330-9Vdiff} we show three nights of $V$-band data for ESO~330-9 with our binary model light curve (described below) subtracted from the
data.  The result shows the variability of the system relative to our binary model.  The width of each window is twice the orbital period.  We only see
the large dimming episode on one night, though the third night does show a small deviation above the uncertainties.  The shift brightward in the third night
may be intrinsic, though the differential photometry zero point is likely only good to within about 0.01 magnitudes based on night-to-night variations, so we
do not take the offset to be significant.
No significant dimming (or brightening) episodes are seen
in the $B$ or $R$ data.  In looking at the episode, the much faster rise time
as opposed to the slower decline suggests that the cause is not a transiting planet or some other solid body, but is more likely a dust cloud, if the cause
is indeed intrinsic to the system.

\begin{figure}[p]
\begin{center}
\includegraphics[width=6in]{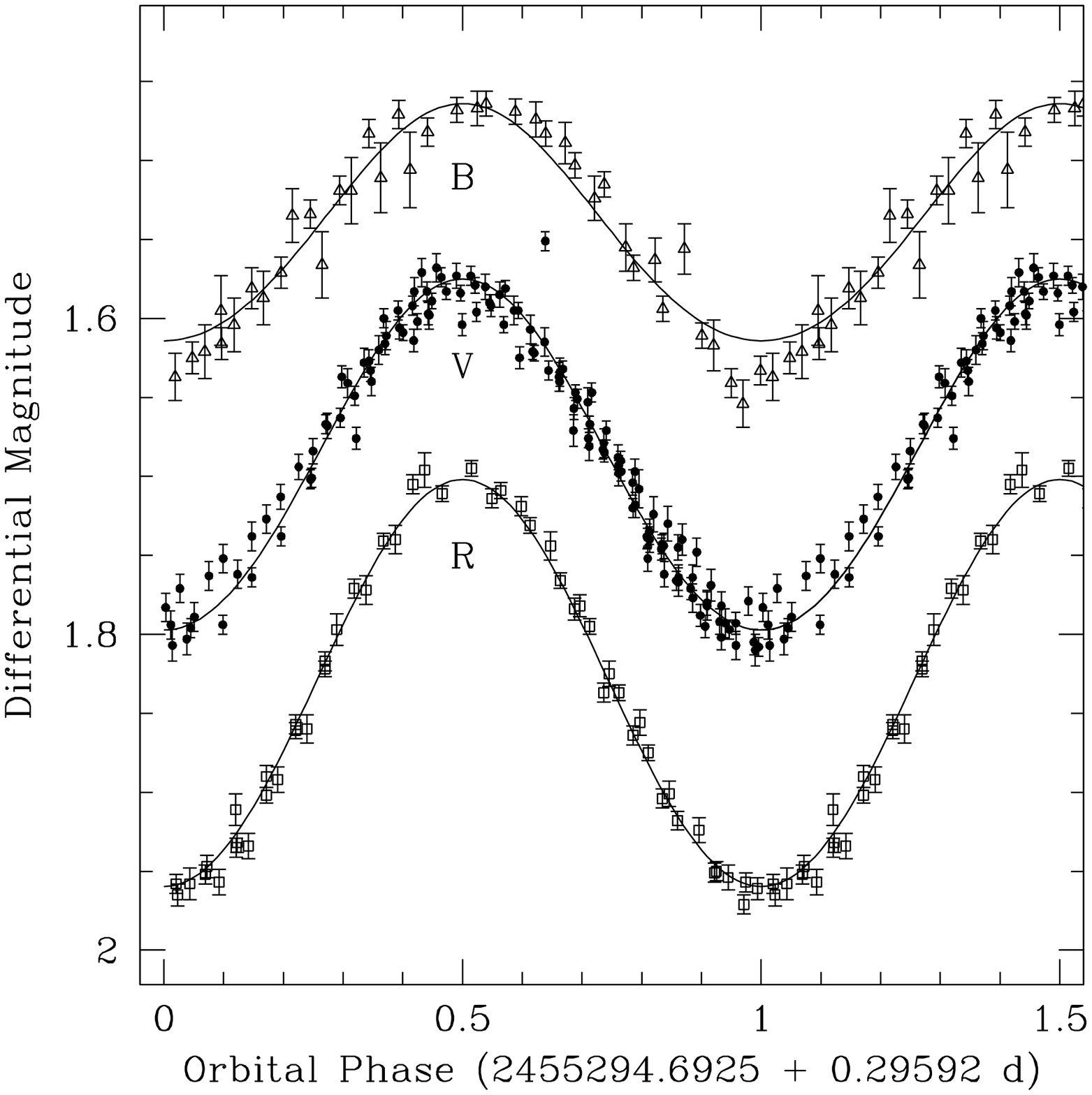}
\end{center}
\caption[ESO~330-9 Phase-Folded Light Curve] {Phased differential magnitude $B$ ({\it triangles}), $V$ ({\it circles}), and $R$ ({\it squares}) light curves
of ESO~330-9 for the ephemeris in the text.  Lines are from Wilson-Devinney
model fits as described in the text.
\label{ESO330-9phot}}
\end{figure}

Because we cannot find a direct relationship between the dimming episode in our $V$-band data and the central binary, for our binary modeling we
have removed that portion of the data so that we can concentrate on fitting the binary system parameters. In Figure \ref{ESO330-9phot} we show the 
resulting three-filter 
light curves for ESO~330-9 along with an example model fit.  The curves have been shifted
vertically for clarity and do not represent relative brightness in the respective filters.  
As with HaTr~7, we see a marked increase in variability amplitude moving from $B$ to $V$ and $R$.

A total of 16 spectra of the CS of ESO~330-9 were obtained with the GMOS instrument at Gemini South in long-slit mode with the B1200 grating with a resolution
of $R=3744$.  The observations
were made in 2013 August with the B1200 grating and a 0.75$\arcsec$ wide slit.  
The spectra cover a range of 4010-5475 \AA~with a spectral
resolution of  0.471 \AA~per pixel.  The IRAF/Gemini package was used to reduce the spectra.  Wavelength calibration used CuAr arc spectra taken consecutively
with the science spectra, giving a typical radial velocity calibration of approximately 0.2 \AA~or better.
\begin{figure}[p]
\begin{center}
\includegraphics[width=4.5in,angle=-90]{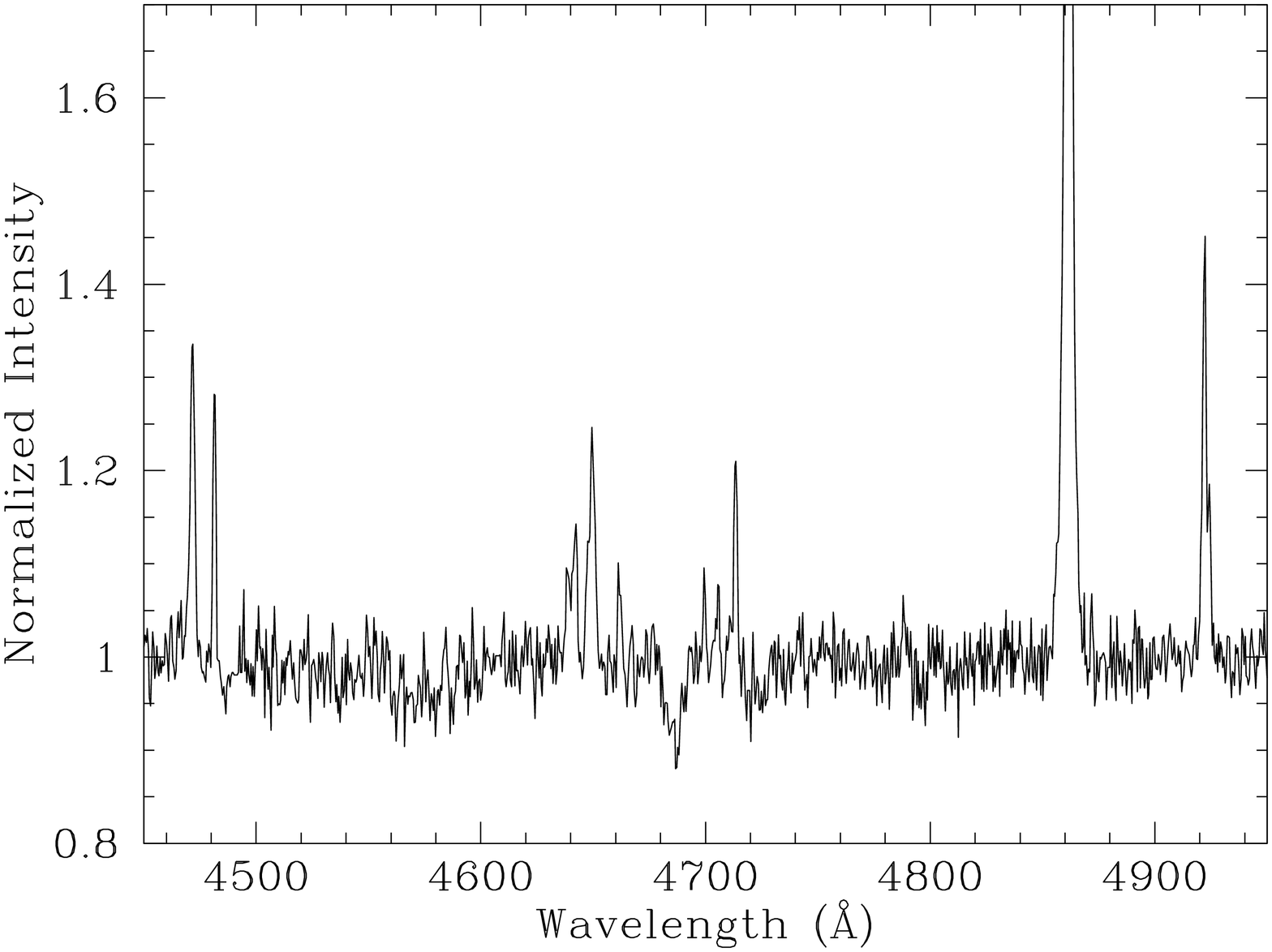}
\end{center}
\caption[ESO~330-9 Spectrum] {A sample spectrum of the CS of ESO~330-9.  The C, N, and O complex of emission lines near 4650 \AA~due to irradiation
of a cool companion is evident as are other lines as described in the text.  Very faint \ion{He}{2} absorption from the hot CS is also present.
\label{ESO330-9spec}}
\end{figure}

The sample spectrum in Figure \ref{ESO330-9spec} shows emission lines from what appears to be an irradiated atmosphere, including the well-recognized C, N, and O
complex near 4650 \AA, Balmer emission, \ion{He}{1} lines at $\lambda\lambda$ 4026, 4388, 4471, and 5015 \AA, \ion{Mg}{2} $\lambda$ 4481 \AA~(partially
overlapped by a chip gap),
and \ion{C}{2} $\lambda$ 4267 \AA.  The appearance of the \ion{He}{1} emission but no \ion{He}{2} suggests a relatively cool CS.  However, the faint and
relatively broad
\ion{He}{2} absorption from the CS also helps to confirm the identity of this object as the CS, even though it appears off-center in the visible nebulosity
(though it is positioned within the high-excitation [\ion{O}{3}] emission).

\begin{figure}[p]
\begin{center}
\includegraphics[width=6in]{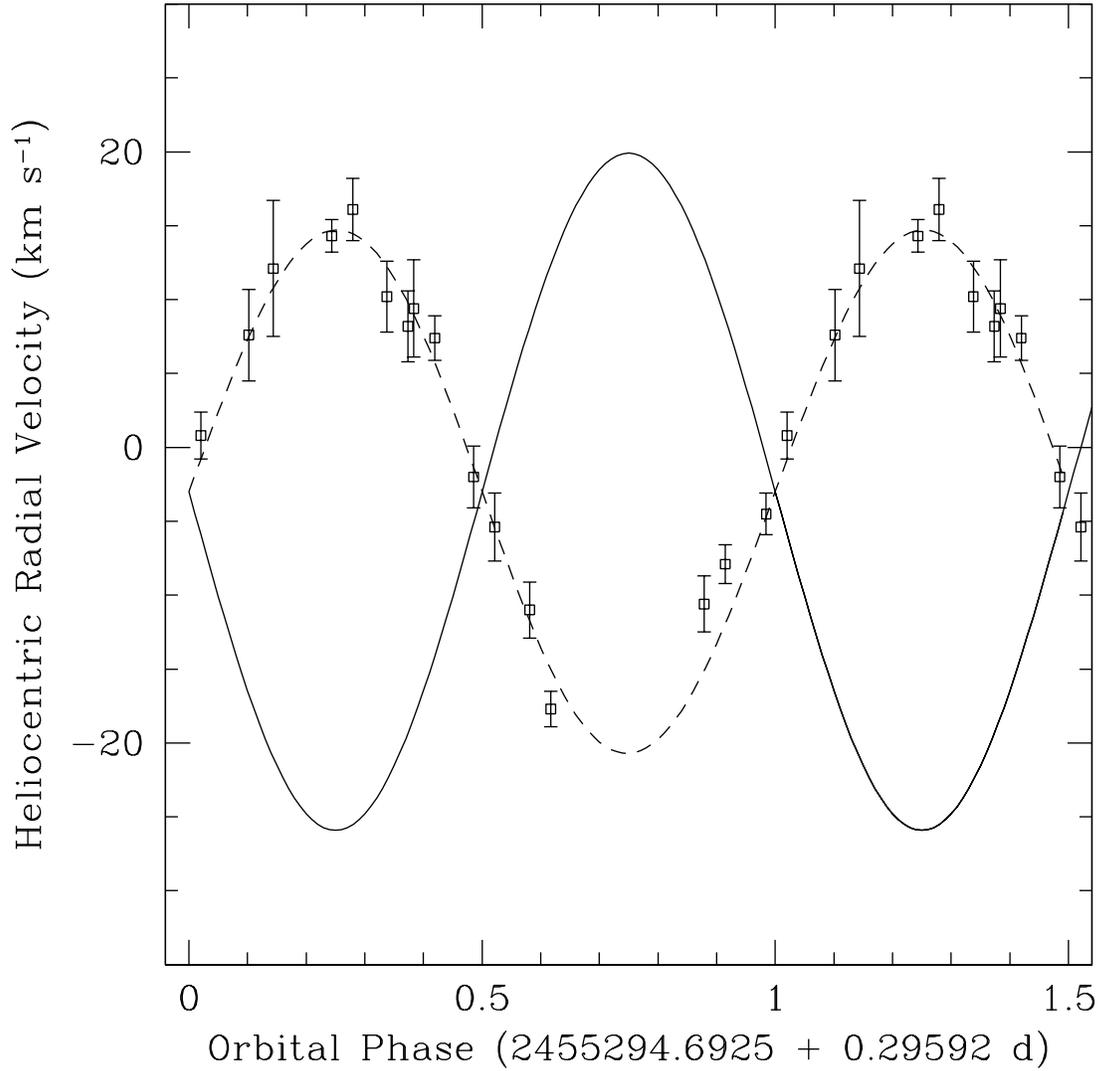}
\end{center}
\caption[Radial velocity curve for the CS of ESO~330-9] {The radial velocity curve for the CS of ESO~330-9.  Data points are for the irradiation lines
described in the text.  The dashed line is the radial velocity curve for the irradiated companion, the solid line is the expected curve for the CS, based on one of
our best-fit parameter sets from binary modeling.
\label{ESO330-9rv}}
\end{figure}

Unfortunately, the \ion{He}{2} absorption is too broad and noisy to give reliable radial velocity measurements at the low radial velocity amplitudes observed here.
A radial velocity curve for the companion is given in Figure \ref{ESO330-9rv}.  The figure shows data points for the emission lines at the photometric period and
one-quarter phase shifted from the photometry, demonstrating that they do occur in the companion
and are due to irradiation.  The dashed and solid lines are example fits for the companion and CS, respectively,
from the binary modeling described below.

\subsection{Binarity and System Parameters}

The low observed radial velocity amplitude of the companion is 15 -- 18 km s$^{-1}$, meaning that either the system is at very low inclination angle, or the
mass ratio, $q=M_2/M_{CS}$ is high.  Given the short orbital period the secondary star cannot be large.  Possible orbital separations for the system are typically
less than 2.5 R$_\odot$.  At inclinations of greater than $i=25^\circ$, in order for the companion star to fit inside its Roche lobe the mass ratio must be
greater than about 3 and results in a CS with a mass below about 0.15 M$_\odot$, which is unlikely, especially given that the CS is ionizing the nebula and is
hot enough to produce the observed \ion{He}{2} absorption line.  To have a CS consistent with evolutionary models and hot enough to ionize the nebula, along
with a companion that fits inside its Roche lobe, we find a maximum inclination of $i\approx15^\circ$.  That requirement also leads to a companion mass
less than $\approx 1$ M$_\odot$.

We keep in mind in this process that the {\it observed}
radial velocity amplitude is the minimum value for the true orbital velocity of the companion.  This is because the curve is determined using the irradiation emission
lines which are produced on the hot hemisphere of the star.  Thus the center of light is closer to the system center of mass, giving lower velocities than for the center
of the star.  The values given above consider the possibility that the companion fills a large portion of its Roche lobe over the various mass ratios and inclinations.

Including the light curves allows us to further constrain the inclination as well as the temperatures and radii of the two stars. 
We find that CS temperatures greater than about 80,000 K cannot reproduce the observed light curves.  Given that our solution
to the binary modeling for the CS of HaTr~7 required changing the limb-darkening parameters, we consider the effect of repeating that for this system.  However,
our observation of low ionization in the irradiation spectrum (the strong \ion{He}{1} and lack of \ion{He}{2}) and the low-excitation nebular spectrum from \citet{fre08}
are both consistent with a relatively cool companion,
so we maintain the internally calculated limb darkening values from the Wilson-Devinney code until we determine if consistent model solutions are possible.

The solutions we find with $T_{CS}$ near 80,000 K result in companion stars with radii significantly smaller than expected for a main sequence star of the
equivalent temperature.  And while companions in these systems are often higher temperature than their MS counterparts, they are also typically larger.  
Thus we are left with lower temperatures for the CS.  We find that
CS temperatures around 55,000 -- 65,000 K give values for temperature and radius for both the CS and companion that are consistent with real stars.  However,
the luminosity we find for the CS is very much smaller than what would be expected for a CO-core post-AGB star.  Comparing our model temperature and
luminosity values to the evolution curves of \citet{dri98} and \citet{hal13}, the CS in our models is consistent with a He-core post-RGB object with masses around
0.38--0.45 M$_\odot$ (see Figure \ref{postAGBRGB}).  Assigning the CS a mass of 0.40 M$_\odot$ and solving for the system values for our best-fit binary
models gives a companion mass of $M_2\approx 0.36$ M$_\odot$ with $R_2\approx0.42$ R$_\odot$ and $T_2\approx 3700$ K.  All of these values are
consistent with an over-luminous MS companion in the system.  We show our best-fit range of binary system values in Table \ref{ESO330-9model}.

While the models in Table \ref{ESO330-9model} provide parameters consistent with physically realistic stars, the evolutionary times present a problem.
For a post-RGB star with $T$ and $\log L/L_\odot$ of our model, the evolutionary time from \citet{dri98} is 102,000 years and about 
98,000 years from \citet{hal13}, significantly longer than the expected lifetime for a PN.  However, the age estimate we calculated above (\S 3.1) is also
long for a PN.  But as we discuss below however, our binary model suggests a distance closer than that of \citet{fre16}, which would give a correspondingly
shorter age of around 45,000 years, roughly half that of the evolutionary age we find for the CS.
It is possible that the CE evolution for this type of system could be accelerated beyond that in the models, but there is no
specific evidence for such an acceleration for ESO~330-9.  \citet{hal13} describe a possible situation in which a post-RGB CS can end the CE
phase in  non-thermal equilibrium, resulting in a faster evolution of the star.  It is also possible that the interaction of the nebula with the surrounding
ISM has delayed the PN evolution by constraining the expanding envelope spatially and/or by collisional reheating.  See section \S 4 
below for further discussion of this topic.

In our discussion to date, we have assumed that ESO 330-9 is a bona fide PN, possibly re-brightened as a result of an ISM interaction \citep[see][]{war10}.  
\citet{par16} class it as a likely PN, but, it is possible that ESO 330-9 might be a mimic as it has an appearance and ionization structure similar to 
K~2-2 \citep{dem13}.  Spatially-resolved, high-resolution spectroscopy of the nebula will allow for the determination of its nature, and if a true PN, 
provide details of the nebular morphology, kinematics, and expansion age.

Using the system brightness values from our model with the measured visual magnitudes, we can calculate a distance to the
PN based on our binary modelling, $d_{model}$.  Given the uncertainty in the $V$ magnitudes and in $E(B-V)$, we use the $I$ and
near-UV magnitudes along with the binary model and $R=3.1$ to determine a new $E(B-V)$ value.  A value of $E(B-V)=0.40$ results in a match
between the observed magnitudes and the model results if the DENIS magnitude was obtained at minimum light.  If the DENIS data gives the
maximum light magnitude, we find $E(B-V)=0.37$.  For the near-UV reddening correction we use the UV extinction relation
described by \citet{val04}.
The resulting distance using the absolute magnitude from our example model (see Table \ref{ESO330-9model} and de-reddened apparent 
magnitude in the $I$ band is about 0.73--0.91 kpc.  
The uncertainty in distance increases for our full range of models from Table \ref{ESO330-9model} 
for which we arrive at $0.7\lesssim d_{model}\lesssim 1.0$ kpc.

Recall that the distance from \citet{fre16} is $1.73\pm0.51$ kpc.  Our range is not in good agreement with their value.  
It is possible that the discrepancy is due to the apparent interaction of the PN with the local ISM.
With the $S_{H\alpha}-r$ relation used by \citet{fre16}, an ISM interaction will usually mean an artificial brightening of the PN, leading to an
{\it underestimated} distance--resulting in a greater discrepancy with our distance.  However, the clear asymmetry of the PN (see Figure \ref{ESO330-9img})
likely also means that the radius is underestimated, bringing the distance back to smaller values.  Another related factor is seen in Figure 5 of \citet{fre16}
where they show some dispersion in the $S_{H\alpha}-r$ relation due to the mass of the CS, where lower mass CSs fall lower in the plot.  Continuing that
relation for the likely mass we find for the CS of ESO~330-9 would lead to a much smaller distance for the same surface brightness.

Without understanding better the ISM interaction in the PN we cannot predict whether our distance and that of \citet{fre16} could be brought into agreement.
However, the dispersion in the relation due to mass does correct in the right direction to bring the two toward better agreement, so we tentatively suggest
that between that consideration and the uncertainties involved in the two distance determinations, our different values of distance do not present an
insurmountable problem for the physical parameters we find from our binary modeling.

\begin{center}
\begin{deluxetable}{lcc}
\tablewidth{0pc}
\tablecolumns{3}
\tablecaption{ESO~330-9 Best-Fit Binary System Parameters
\label{ESO330-9model}}
\tablehead{
\colhead {Parameter}        & \colhead{Range}    & \colhead{Example\tablenotemark{a}}}
\startdata
$T_{CS}$ (kK) 		& 55--65			&  60 \\
$M_{CS}$ (M$_\odot)$	& 0.38--0.45 	&  0.40 \\
$T_2$ (K) 		& $\lesssim 4500$	&  3700 \\
$R_{CS}$ 		& 0.03--0.07 		&  0.040 \\
$M_2$ (M$_\odot)$	& 0.3--0.5		 	&  0.36 \\
$R_2$			& 0.35--0.50	 	&  0.42 \\
$a$ (R$_\odot$)	& 1.6 -- 1.8  		&  1.71 \\
$q$				& 0.8 -- 1.2	 	&  0.90 \\
$i$ ($^\circ$) 		& 7 -- 13 			&  9.5 \\
\enddata
\tablenotetext{a}{See text for details}
\end{deluxetable}
\end{center}

\section{Post-RGB evolution and PNe}

Observational evidence for multiple evolutionary pathways to planetary nebula formation is suggestive
but not yet unequivocal \citep{fre10b, fre14, rei14}.  No reliable
candidates for post-RGB planetary nebulae were identified by \citet{hal13}, while a candidate
for a post-EHB nebula, PHL 932, was shown by \citet{fre10b} to be ionized interstellar matter.
On the other hand, \citet{gei11} propose that the similar nebula EGB~5 might be the result of a CE ejection.
Since then, \citet{hil16a} identified HaTr\,4 and Hf\,2-2 as candidate close-binary
initiated post-RGB nebulae.  Also, \citet{jon16} have identified Abell\,46 as another possible
post-RGB PN.  The hot CS in Abell\,46, V477~Lyr, has a measured mass of $0.508\pm 0.046$ M$_\odot$ \citep{afs08},
which is borderline between post-AGB and post-RGB evolution.  
Their low temperature of $49500\pm4500$ K
is also consistent with a post-RGB object.  However, \citet{shi08} give $T=83000\pm5000$ K, though their
reported mass value for the CS of $0.556\pm 0.096$ M$_\odot$ could still allow for a post-RGB CS.
However, \citet{jon16} also
note that post-RGB evolution may be mechanism through which high and extreme abundance
discrepancy factors could be explained, with Abell\,46 having a high ADF value.

In Table~\ref{tab:post-RGB}, we summarize the five objects currently
considered as possible post-RGB planetary nebulae, including the CS of ESO 330-9, along with three
hot stars with similar properties, but which are currently thought to be ionizing regions of ambient ISM
\citep{fre10a, fre14, dem13} though \citet{ran10, ran15} offer a contrary opinion.

Some other PNe which may host low-luminosity CSs as inferred from the distances and flux data in \citet{fre16} and \citet{par16}
are Hen~2-105 \citep{wei11}, Sp~3 \citep{gau01}, NGC~6026 \citep{hil10, dan13}, 
K~1-2 \citep{ext03}, RWT~152 \citep{all16}, SkAc~1 \citep{dou15}, 
and M~3-57 \citep[= HaWe 11;][]{har87}.  
Further work is needed to place them in the HR diagram, before they can be considered as candidate post-RGB PNe.

Recent models of CE interactions  \citep*{dem11,nie12,mad16}
predict that as many post-RGB as post-CE close binary stars may be produced.  Those studies typically
do not include post-RGB systems as PNe since typical evolutionary times \citep{dri98,hal13} are much longer
than predicted PN lifetimes.  However, \cite{hal13} describe possible cases in which post-RGB
systems may produce a visible PN, which they suggest can occur for CS masses greater than about 0.26 M$_\odot$.  
It is not clear from their work what fraction of CSs greater than this mass may form a visible PN, but most populations synthesis
and CE evolution models predict that a majority of post-RGB objects will have masses in this range.
There is a possibility then that a significant fraction of post-RGB objects may produce a visible PN, and that this fraction
may even approach the fraction of PNe with post-AGB close binaries.
The statistics to date for CSPNe \citep[e.g.][]{hil14} show far fewer post-RGB objects currently known than 
post-AGB close binaries.  However, the distances to many close-binary PNe
are not very reliable, so it is possible that some systems currently thought to be post-AGB objects may
result instead from post-RGB evolution. 
Furthermore, \citet{kam16} has recently found a population of dusty post-RGB stars in the LMC that may 
have formed as a result of binary interaction. Whether these objects are the precursors of post-RGB PNe remains to be determined.  
Eventually accurate distances from the GAIA survey \citep{gai16}
will allow the luminosities of each system to be reliably measured, allowing for
their exact evolutionary origins to be determined, and allowing for a comparison with population-synthesis
models.  Obtaining orbit-resolved spectra for more of these close binary CSPNe will also allow more detailed
binary modeling which can produce CS masses, providing a further test of evolutionary channels.

\begin{table*}
\begin{center}
\caption{Candidate post-RGB PNe}\label{tab:post-RGB}
\begin{tabular}{llccll}
\hline
Name		&	~~~~~~$T_{\rm eff}$	&	     log\,$g$	   	& Status$^{\ast}$	& References		&      	Notes			\\
\hline
HaTr 4	  	&	60000 $\pm$ 10000	&		...		  	&	T			& HB16			&	...				\\
Hf 2-2		&	79000 $\pm$ 8000	&		...			&	T			& HB16			&	High ADF			\\
Abell 46		&	49500 $\pm$ 4500	&      	5.67 $\pm$ 0.05	&	T			& PB94, AI08		&	High ADF			\\
ESO 330-9	&	60000 $\pm$ 5000 	&		...			&	L			& This work		&	...				\\	
HaWe 13		&	68100 $\pm$ 9400 	&	6.38 $\pm$ 0.31 	&	P			& HW87, N99, FP16	&	subluminous PN?	\\
\hline
K 2-2		&	67000 $\pm$ 11000 	&	6.09 $\pm$ 0.24	&	N			& N99, DM13		&	mimic			\\
Sh 2-174		&	69100 $\pm$ 3000 	&	6.70 $\pm$ 0.18 	&	N			& N99, FP10	&	mimic			\\
DHW 5		&	76500 $\pm$ 5800 	&	6.65 $\pm$ 0.19	&	N			& N99, FP10	&	mimic			\\
\hline
\end{tabular}
\end{center}
\begin{flushleft}
References:~~AI08 -- \citet{afs08}; DM13 --  \citet{dem13}; FP10 --  \citet{fre10b}; FP16 --  \citep{fre16}; HB16 -- \citet{hil16a}; HW87 -- \citet{har87};
N99 --  \citet{nap99}; PB94 -- \citet{pol94}; $^{\ast}$ \citet{par16}.
\end{flushleft}

\end{table*}

\section{Summary}

The CSs of both HaTr~7 and ESO~330-9 show photometric variability due to an irradiated cool companion.  The variability period for each system is equal to
the orbital period determined from radial velocity curves.  For HaTr~7, the binary CS is a double-lined spectroscopic binary showing emission lines from the
irradiated hemisphere of  the companion and both hydrogen Balmer and \ion{He}{2} absorption lines.  For the CS of ESO~330-9 we have a single-lined radial
velocity curve based on emission lines from the companion.

We have determined orbital periods along with producing limits on the physical parameters for each system using binary modeling.  In the case of HaTr~7
the double-lined radial velocities along with modeling of the spectral lines provided much tighter constraints on the system parameters.  For that system, our CS parameters
from spectral modeling and binary modeling are in agreement with one another, though we find that in order to properly reproduce the light curves we need to modify
the limb darkening coefficients for the companion star.

The CS of ESO~330-9 however, is only consistent with theoretical evolutionary tracks if the star is a He-core, likely post-RGB object.  The temperature and luminosity
of the CS required to fit the light curves corresponds to a well-evolved object with mass $\approx0.4$ M$_\odot$ that is significantly older than would be expected for
a CSPN.  Though it is possible that the evolution of the CS was accelerated due to the binary interaction and possible non-thermal equilibrium processes.  The
possibility of a post-RGB close-binary CSPN would be an interesting test of existing models of PN evolution as well as the formation of close binaries through CE evolution.
More detailed studies of the nebular expansion rate, ionization, and composition would help to clarify the picture.

Using our parameter ranges we calculate a distance to each PN.  For HaTr~7 the distance range is consistent with distances found previously, though for
ESO~330-9 the agreement is marginal, with our results showing a system that is significantly closer.

ESO~330-9 is a diffuse PN that clearly seems to be interacting with the surrounding ISM.  As such it does not have any clear large-scale structure or symmetries
that would allow determination of an inclination that could be matched to the inclination of the binary system from our modeling.  However, HaTr~7 shows a symmetric
structure that, though it is faint, could potentially be modeled to find an inclination.  This could then be compared to our binary system inclination and added to the
collection of systems for which both are known \citep[see][]{hil16b}.

Apart from comparing inclinations, we have added two more systems to the growing number of close binary CSPNe.  As the sample grows we can make increasingly
statistically significant studies of close binary CSPNe.  Those studies will be able to tell us more about CE evolution, PN formation and ejection, and potentially
the formation of cataclysmic variables and type-Ia supernova progenitors.

\acknowledgements

The authors would like to thank David Jones for helpful comments on the text.
This material is based upon work supported by the National Science Foundation under Grant No. AST-1109683. Any opinions, findings, and conclusions or recommendations expressed in this material are those of the author(s) and do not necessarily reflect the views of the National Science Foundation.  
Based on observations obtained with the SARA Observatory 0.6 m telescope at Cerro Tololo, which is operated by the Southeastern Association for Research in Astronomy (saraobservatory.org).
Based on observations (GS-2012A-Q-85 and GS-2015A-Q-97) obtained at the Gemini Observatory, processed using the Gemini IRAF package, which is operated by the Association of Universities for Research in Astronomy, Inc., under a cooperative agreement with the NSF on behalf of the Gemini partnership: the National Science Foundation (United States), the National Research Council (Canada), CONICYT (Chile), Ministerio de Ciencia, Tecnología e Innovación Productiva (Argentina), and Ministério da Ciência, Tecnologia e Inovação (Brazil).
The CSS survey is funded by the National Aeronautics and Space
Administration under Grant No. NNG05GF22G issued through the Science
Mission Directorate Near-Earth Objects Observations Program.  The CRTS
survey is supported by the U.S.~National Science Foundation under
grants AST-0909182 and AST-1313422.

{\it Facility:} \facility{Gemini: South}, \facility{SOAR}, \facility{SARA:CTIO}


\end{document}